\documentclass[12pt]{article}
\usepackage{amsmath,color}
\usepackage[dvips]{graphicx}
\input amssym
\input amssym.def

\def\red#1{{\color{red} #1}}
\begin{document}

\def\prg#1{\medskip\noindent{\bf #1}}  \def\ra{\rightarrow}
\def\lra{\leftrightarrow}              \def\Ra{\Rightarrow}
\def\nin{\noindent}                    \def\pd{\partial}
\def\dis{\displaystyle}                \def\inn{\hook}
\def\grl{{GR$_\Lambda$}}               \def\Lra{{\Leftrightarrow}}
\def\cs{{\scriptstyle\rm CS}}          \def\ads3{{\rm AdS$_3$}}
\def\Leff{\hbox{$\mit\L_{\hspace{.6pt}\rm eff}\,$}}
\def\bull{\raise.25ex\hbox{\vrule height.8ex width.8ex}}
\def\ric{{Ric}}                      \def\tric{{(\widetilde{Ric})}}
\def\Lie{{\cal L}\hspace{-.7em}\raise.25ex\hbox{--}\hspace{.2em}}
\def\sS{\hspace{2pt}S\hspace{-0.83em}\diagup}   \def\hd{{^\star}}
\def\dis{\displaystyle}                 \def\ul#1{\underline{#1}}
\def\mb#1{\hbox{{\boldmath $#1$}}}     \def\tgr{{GR$_\parallel$}}
\def\irr#1{^{(#1)}\hspace{-2pt}}

\def\hook{\hbox{\vrule height0pt width4pt depth0.3pt
\vrule height7pt width0.3pt depth0.3pt
\vrule height0pt width2pt depth0pt}\hspace{0.8pt}}
\def\semidirect{\;{\rlap{$\supset$}\times}\;}
\def\first{\rm (1ST)}       \def\second{\hspace{-1cm}\rm (2ND)}
\def\bm#1{\hbox{{\boldmath $#1$}}}
\def\nb#1{\marginpar{{\large\bf #1}}}
\def\ir#1{{}^{(#1)}}  \def\ArcTan{\text{ArcTan}}

\def\G{\Gamma}        \def\S{\Sigma}        \def\L{{\mit\Lambda}}
\def\D{\Delta}        \def\Th{\Theta}
\def\a{\alpha}        \def\b{\beta}         \def\g{\gamma}
\def\d{\delta}        \def\m{\mu}           \def\n{\nu}
\def\th{\theta}       \def\k{\kappa}        \def\l{\lambda}
\def\vphi{\varphi}    \def\ve{\varepsilon}  \def\p{\pi}
\def\r{\rho}          \def\Om{\Omega}       \def\om{\omega}
\def\s{\sigma}        \def\t{\tau}          \def\eps{\epsilon}
\def\nab{\nabla}      \def\btz{{\rm BTZ}}   \def\heps{\hat\eps}
\def\bt{{\bar t}}     \def\br{{\bar r}}    \def\bth{{\bar\theta}}
\def\bvphi{{\bar\vphi}}
\def\bx{{\bar x}}     \def\by{{\bar y}}     \def\bom{{\bar\om}}
\def\tphi{{\tilde\vphi}}  \def\tt{{\tilde t}}

\def\tG{{\tilde G}}   \def\cF{{\cal F}}      \def\bH{{\bar H}}
\def\cL{{\cal L}}     \def\cM{{\cal M }}     \def\cE{{\cal E}}
\def\cH{{\cal H}}     \def\cA{{\cal A}}      \def\hcH{\hat{\cH}}
\def\cK{{\cal K}}     \def\hcK{\hat{\cK}}    \def\cT{{\cal T}}
\def\cO{{\cal O}}     \def\hcO{\hat{\cal O}} \def\cV{{\cal V}}
\def\tom{{\tilde\omega}}                     \def\cE{{\cal E}}
\def\cR{{\cal R}}    \def\hR{{\hat R}{}}     \def\hL{{\hat\L}}
\def\tb{{\tilde b}}  \def\tA{{\tilde A}}     \def\tv{{\tilde v}}
\def\tT{{\tilde T}}  \def\tR{{\tilde R}}     \def\tcL{{\tilde\cL}}
\def\hy{{\hat y}\hspace{1pt}}  \def\tcO{{\tilde\cO}}
\def\bA{{\bar A}}     \def\bB{{\bar B}}      \def\bC{{\bar C}}
\def\bG{{\bar G}}     \def\bD{{\bar D}}      \def\bH{{\bar H}}
\def\bK{{\bar K}}     \def\bL{{\bar L}}

\def\rdc#1{\hfill\hbox{{\small\texttt{reduce: #1}}}}
\def\chm{\checkmark}  \def\chmr{\red{\chm}}
\def\nn{\nonumber}                    \def\vsm{\vspace{-3pt}}
\def\be{\begin{equation}}             \def\ee{\end{equation}}
\def\ba#1{\begin{array}{#1}}          \def\ea{\end{array}}
\def\bea{\begin{eqnarray} }           \def\eea{\end{eqnarray} }
\def\beann{\begin{eqnarray*} }        \def\eeann{\end{eqnarray*} }
\def\beal{\begin{eqalign}}            \def\eeal{\end{eqalign}}
\def\lab#1{\label{eq:#1}}             \def\eq#1{(\ref{eq:#1})}
\def\bsubeq{\begin{subequations}}     \def\esubeq{\end{subequations}}
\def\bitem{\begin{itemize}}           \def\eitem{\end{itemize}}
\renewcommand{\theequation}{\thesection.\arabic{equation}}
\title{Hamiltonian approach to black hole entropy:\\ Kerr-like spacetimes}

\author{M. Blagojevi\'c and B. Cvetkovi\'c\footnote{
        Email addresses: \texttt{mb@ipb.ac.rs, cbranislav@ipb.ac.rs}} \\
Institute of Physics, University of Belgrade,\\
                      Pregrevica 118, 11080 Belgrade-Zemun, Serbia}
\date{\today}
\maketitle

\begin{abstract}
Black hole entropy of the Kerr-like family of spacetimes is introduced as the canonical charge on the horizon. Treating these spacetimes in the framework of Poincar\'e gauge theory either as Riemannian solutions or as  solutions with torsion, it is shown that these geometrically different setups lead to the same black hole entropy.
\end{abstract}
\section{Introduction}
\setcounter{equation}{0}

In general relativity (GR), black hole entropy can be classically interpreted as the Noether charge on horizon \cite{x1,x2}. In an attempt to explore the notion of entropy in Poincar\'e gauge theory (PG) \cite{x3,x4,x5}, where both the torsion $T^i$ and the curvature $R^{ij}$ define the gravitational dynamics, we used a Hamiltonian formulation of this idea to introduce a generalized concept of black holes entropy \cite{x6}. The result was successfully applied to several exact solutions of PG, belonging to the family of spherically symmetric solutions with or without torsion.

Kerr spacetime \cite{x7,x8,x9,x10,x11,x12} is a stationary and axially symmetric solution of GR with a vanishing effective cosmological constant $\l$,  which has played an important role in understanding dynamical features of black holes in GR \cite{x2}. In the present paper, we extend our Hamiltonian approach to the study of entropy of Kerr-like spacetime, the spacetime with Kerr metric which is either a \emph{torsionless} or a solution \emph{with torsion} of the general (parity preserving) PG, including the teleparallel theory of gravity (TG) as its subcase. The analysis of these two types of solutions clarifies how different geometric setups, characterized by the absence or presence of torsion, affect black hole entropy, and moreover, it offers a deeper insight into the Hamiltonian description of black hole entropy.

The paper is organized as follows. In section \ref{sec2}, we review basic aspects of PG and the Hamiltonian approach to black hole entropy. In section \ref{sec3}, we introduce the tetrad formulation of the Kerr geometry, needed in the Hamiltonian treatment of entropy. In section \ref{sec4}, we use this formalism to study entropy (and conserved charges) of the Kerr black hole treated as a Riemannian solution of PG. In section \ref{sec5}, the analysis is extended to Kerr black holes with torsion in the framework of PG and its subcase TG. Finally, section \ref{sec6} is devoted to a discussion of the results.

Our conventions are the same as in Ref. \cite{x6}. Latin indices $(i,j,\dots)$ refer to the local Lorentz frame, greek indices $(\m,\n,\dots)$ refer to the coordinate frame,  $b^i$ is the orthonormal coframe (tetrad 1-form), $h_i$ is the dual basis (frame) such that $h_i\inn b^k = \d_i^k$, and the Lorentz metric is $\eta_{ij}=(1,-1,-1,-1)$. The volume 4-form is
$\heps = b^0∧ b^1∧ b^2∧b^3$, the Hodge dual of a form $\a$ is denoted by $\hd\a$, with $\hd 1=\heps$, and the totally antisymmetric symbol $\ve_{ijmn}$ is normalized to $\ve_{0123}=+1$. The exterior product of forms is implicit.

\section{General formalism}\label{sec2}
\setcounter{equation}{0}

We begin by giving a short account of PG and reviewing basic aspects of the Hamiltonian  approach to black hole entropy introduced in Ref. \cite{x6}.

Basic dynamical variables of PG are the tetrad field $b^i$ and the antisymmetric spin connection $\om^{ij}$ (1-forms), the corresponding field strengths are the torsion $T^i=d b^i+\om^i{_k} b^k$ and the curvature $R^{ij}=d\om^{ij}+\om^i{_k}\om^{kj}$ (2-forms), and the underlying structure of  spacetime is described by a Riemann-Cartan geometry \cite{x3,x4,x5}.
In the absence of matter, gravitational dynamics of PG is determined by the gravitational Lagrangian $L_G(b^i,T^i,R^{ij})$ (4-form), which is assumed to be at most quadratic in the field strengths and parity invariant,
\be
L_G=-\hd(a_0R+2\L)+T^i\sum_{n=1}^3\hd(a_n\ir{n}T_i)
            +\frac{1}{2}R^{ij}\sum_{n=1}^6\hd(b_n\ir{n}R_{ij})\,, \lab{2.1}
\ee
where $(a_0,\L,a_n,b_n)$ are the coupling constants, and $\ir{n}T_i,\ir{n}R_{ij}$ are irreducible parts of the field strengths, see, for instance, Ref. \cite{x6}. The gravitational field equations in vacuum are obtained by varying $L_G$ with respect to $b^i$ and $\om^{ij}$. After introdu\-cing the covariant momenta, $H_i=\pd L_G/\pd T^i$ and $H_{ij}=\pd L_G/\pd R^{ij}$, and the corresponding energy-momentum and spin currents, $E_i=\pd L_G/\pd b^i$ and $E_{ij}=\pd L_G/\pd\om^{ij}$, these equations can be written in a compact form as \cite{x4,x5}
\bsubeq\lab{2.2}
\bea
\d b^i:&&\nab H_i+E_i=0\,,                                      \lab{2.2a}\\
\d\om^{ij}:&&\nab H_{ij}+E_{ij}=0\,.                            \lab{2.2b}
\eea
\esubeq

Recently \cite{x6}, we proposed a general Hamiltonian approach to black hole entropy in PG. The approach is based on the existence of the canonical gauge generator $G$, defined as an integral over the spatial section $\S$ of spacetime. Since $G$ acts on the phase-space variables via the Poisson bracket operation, it should have well-defined functional derivatives.
This property can be ensured by adding to $G$ a suitable surface term $\G$, an integral over the boundary of $\S$, such that the improved generator $\tG:=G+\G$ is well defined (or regular),
\be
\d\tG\equiv \d(G+\G)=R\, ,                                       \lab{2.3}
\ee
where $R$ stands for regular.

In case of a black hole spacetime, we assume that the boundary of $\S$ has \emph{two components}, one at infinity and the other at the horizon, $\pd\S=S_\infty\cup S_H$. As a consequence, the boundary term $\G$ has two parts, $\G:=\G_\infty-\G_H$ (the minus sign in front of $\G_H$ reflects the change in orientation), which are defined by the variational equations \cite{x6}
\bsubeq\lab{2.4}
\bea
&&\d\G_\infty=\oint_{S_\infty}\d B(\xi)\,,\qquad
       \d\G_H=\oint_{S_H} \d B(\xi)\,,                                  \\
&&\d B(\xi):=(\xi\inn b^{i})\d H_i+\d b^i(\xi\inn H_i)
   +\frac{1}{2}(\xi\inn\om^{ij})\d H_{ij}
   +\frac{1}{2}\d\om^{ij}(\xi\inn\d H_{ij})\, .
\eea
\esubeq
Here, symmetries of a Kerr spacetime are described by the Killing vectors $\xi=\pd_t$ and $\pd_\vphi$, whereas the Lorentz parameter $\th^{ij}$ is absent as its presence would not be compatible with the symmetries defined by Killing vectors $\xi$, as explained in section \ref{sec4}.  The variational procedure is assumed to satisfy the following requirements:
\bitem
\item[a1)] the variation $\d\G_\infty$ is performed over a suitable set of asymptotic states, leaving the background configuration fixed;\vsm
\item[a2)] the variation $\d\G_H$ is performed by varying characteristic parameters of a solution, but keeping surface gravity constant (the zeroth law of black hole thermodynamics).
\eitem
The choice of asymptotic conditions for dynamical variables is based on two criteria: first, it should be sufficiently general to include the solution for which the charges are to be calculated, and second, it should be sufficiently restricted so as to ensure finiteness of the corresponding integrals.
When $\G_\infty$ and $\G_H$ are solutions of the variational equations \eq{2.4}, $\G_\infty$ is interpreted as the \emph{asymptotic charge} (energy-momentum or angular momentum), whereas $\G_H$ defines \emph{entropy} as the canonical charge on the horizon.

In general, the explicit form of the entropy term $\G_H$ depends on dynamical properties of the theory and the specific structure of the black hole. For stationary black holes in GR, the entropy formula takes the standard form
\be
\d\G_H=T\d S\, ,                                                   \lab{2.5}
\ee
where $T=\k/2\pi$ is the temperature and $S$ is black hole entropy, given by the area law.

Returning to the regularity condition \eq{2.3}, one should note that it can be expressed in an equivalent form as
\be
\d G=-\d\G+R\, .
\ee
This formula reveals an important result: the gauge generator $G$ is regular if and only if
\be
\d\G\equiv \d\G_\infty-\d\G_H=0\, ,                                \lab{2.7}
\ee
which represents the well-known \emph{first law} of black hole thermodynamics.
Thus:
\bitem
\item[a3)] regularity of the generator $G$ is equivalent to the validity of the first law.
\eitem
In what follows, the results of the present approach will be used to study dynamical origin of entropy and the first law in Kerr-like spacetimes.

\section{Tetrad form of Kerr geometry}\label{sec3}
\setcounter{equation}{0}

We devote this section to general aspects of the tetrad formulation of Kerr geometry, which is needed for the Hamiltonian analysis of entropy \cite{x6}.

Kerr metric in Boyer-Lindquist coordinates can be written in the form \cite{x11}
\bsubeq\lab{3.1}
\be
ds^2=N^2\Big(dt+a\sin^2\th d\vphi\Big)^2-\frac{dr^2}{N^2}
  -\r^2d\th^2-\frac{\sin^2\th}{\r^2}\Big[a dt+(r^2+a^2)d\vphi\Big]^2\,,
\ee
where
\be
N^2:=\frac{\D}{\r^2}\, ,\qquad
\D:=r^2+a^2-2mr\,,\qquad\r^2:=r^2+a^2\cos^2\th\, .
\ee
\esubeq
Since the metric is not singular at $r=0$, $r$ is allowed to run over the whole real line. Thus, $t$ and $r$ belong to the plane $R^2$, $\vphi$ and $\th$ are on the 2-sphere $S^2$, and spacetime is the product manifold $R^2\times S^2$.
The metric \eq{3.1} is singular at $\D=0$ (horizon) and $\r^2=0$ (the ring singularity), see O'Neill \cite{x9}.

The Kerr metric has two Killing vectors, $\pd_t$ and $\pd_\vphi$. For $m=0$, it  reduces to the Minkowski form and for $a=0$ it takes the Schwarzschild form.  The outer horizon is located at the larger root $r=r_+$ of $\D=0$,
\be
2mr_+=r_+^2+a^2\, ,                                               \lab{3.2}
\ee
and the quantity
\be
\Om_+=\frac{g_{t\vphi}}{g_{\vphi\vphi}}\Big|_{r_+}
     =\frac{a}{2mr_+}=\frac{a}{r_+^2+a^2}
\ee
is the angular velocity of the horizon with respect to the Minkowski background.

The form of $\Om_+$ ensures that the Killing vector $\xi=\pd_t-\Om_+\pd_\vphi$ is a null vector on the horizon, $\xi^2=0$, and normal to it. The surface gravity $\k$ is defined by the relation $\pd_\m\xi^2=-2\k\xi_\m$. To calculate it, one should make a transformation from the Boyer-Lindquist coordinates, which are singular at the horizon, to the Edington-Finkelstein-like coordinates \cite{x10}. The final result is
\be
\k=\frac{r_+-m}{2mr_+}\, .
\ee

The form of the metric \eq{3.1} suggests the following choice for the orthonormal tetrad:
\bea
&&b^0=N\Big(dt+a\sin^2\th\,d\vphi\Big)\,,\qquad b^1=\frac{dr}{N}\,,\nn\\
&&b^2=\r d\th\, ,\qquad
  b^3=\frac{\sin\th}{\r}\Big[a\,dt+(r^2+a^2)d\vphi\Big]\, .       \lab{3.5}
\eea
The area of the horizon is given by
\be
A=\int_{S_H}b^2b^3=4\pi (r_+^2+a^2)\, .
\ee

Riemannian connection $\tom^{ij}$ is defined by the condition of vanishing torsion:
\bea\lab{3.7}
&&\tom^{01}=-N'b^0-\frac{ar}{\r^3}\sin\th b^3\,,\qquad
  \tom^{02}=\frac{a^2\sin\th\cos\th}{\r^3}b^0
                                 -\frac{aN}{\r^2}\cos\th b^3\,,   \nn\\
&&\tom^{03}=-\frac{ar}{\r^3}\sin\th b^1+\frac{aN}{\r^2}\cos\th b^2\,,
                                                                  \nn\\
&&\tom^{12}=\frac{a^2\sin\th\cos\th}{\r^3}b^1+\frac{Nr}{\r^2}b^2\,,
  \qquad
  \tom^{13}=-\frac{ar}{\r^3}\sin\th b^0+\frac{Nr}{\r^2}b^3\,,     \nn\\
&&\tom^{23}=
   -\frac{aN}{\r^2}\cos\th b^0+\frac{(a^2+r^2)\cos\th}{\r^3\sin\th}b^3\,.
\eea

The Riemannian curvature $R^{ij}=d\tom^{ij}+\tom^i{}_k\tom^{kj}$ has a rather complicated form:
\bsubeq\lab{3.8}
\bea
&&R_{01}=2Cb^0b^1+2D b^2b^3\,,\qquad R_{23}=-2Cb^2b^3+2D b^0b^1\,,\nn\\
&&R_{02}=-Cb^0b^2+Db^1b^3\, ,\qquad  R_{03}=-Cb^0b^3 -D b^1b^2\,, \nn\\
&&R_{12}= Cb^1b^2-Db^0b^3\, ,\qquad  R_{13}= Cb^1b^3 +D b^0b^2\, ,
\eea
where
\be
C:=\frac{mr}{\r^6}(r^2-3a^2\cos^2\th)\, ,\qquad
D:=\frac{ma\cos\th}{\r^6}(3r^2-a^2\cos^2\th)\, .
\ee
\esubeq
It has only one nonvanishing irreducible part, $\irr{1}R^{ij}=R^{ij}$. For  $m=0$, the curvature vanishes and Kerr spacetime reduces to the Minkowski form.
The quadratic curvature invariant,
\be
R^{ij}\hd R_{ij}=\frac{24 m^2}{\r^{12}}\big(r^2-a^2\cos^2\th\big)
                          \big(\r^4-16a^2r^2\cos^2\th\big)\heps\,,  \lab{3.9}
\ee
exhibits a singularity at $\r^2=0$ (shaped as a ring $r=0,\th=\pi/2$), but not at
the horizon $r=r_\pm$. In fact, a Kerr metric with real $r_\pm$ can be extended from the exterior region $r>r_+$ across both horizons into $r<r_-$; for more details, see Refs. \cite{x9,x10,x12}.

The above considerations can be extended to include the presence of torsion, transforming thereby the Riemanian geometry of spacetime into a Riemann-Cartan form.

\section{Riemannian Kerr black hole in PG}\label{sec4}
\setcounter{equation}{0}

Now, we wish to study the Kerr black hole as a torsionless spacetime in the  framework of PG. Since the Kerr black hole is Ricci flat, $\ric^k=h_i\inn R^{ik}=0$, it satisfies the GR field equations with a vanishing cosmological constant, $\L=0$. Relying on the result that all solutions of GR are also solutions of PG, see Obukhov \cite{x5}, one can conclude that the Kerr black hole satisfies also the PG field equations for $\L=0$. A direct proof is based on the form of the effective PG Lagrangian,
\be
L_G=-\hd(a_0 R+2\L)+\frac{1}{2}b_1 R^{ij} \hd R_{ij}\, ,           \lab{4.1}
\ee
in combination with the field equations \eq{2.2}.
The effective Lagrangian defines the corresponding covariant momenta as $H_i=0$ and
\bsubeq
\be
H_{ij}=-2a_0\hd(b_ib_j)+b_1\hd R_{ij}\, ,
\ee
or, in more detail,
\bea
&&H_{01}=-2a_0b^2b^3+2b_1(2Cb^2b^3+2Db^0b^1)\, ,                   \nn\\
&&H_{02}=~~2a_0b^1b^3+2b_1(Cb^1b^3-Db^0b^2)\, ,                    \nn\\
&&H_{03}=-2a_0b^1b^2+2b_1(-Cb^1b^2-Db^0b^3)\, ,                    \nn\\
&&H_{12}=-2a_0b^0b^3+2b_1(Cb^0b^3-Db^1b^2)\,,                      \nn\\
&&H_{13}=~~2a_0b^0b^2+2b_1(-Cb^0b^2-Db^1b^3)\,,                    \nn\\
&&H_{23}=-2a_0b^0b^1+2b_1(-2Cb^0b^1+2Db^2b^3)\,.
\eea
\esubeq

Basic dynamical variables of the Kerr geometry are invariant under the action of the Killing vectors $\xi=\pd_t$ and $\pd_\vphi$, $\d_0 b^i{_\m}=0$ and $\d_0\om^{ij}{}_\m=0$ (no $t$ and/or $\vphi$ dependence). In particular, this property is not compatible with any Lorentz transformation, so that the corresponding parameter $\th^{ij}$ has to be exactly zero. For Riemannian solutions, the variational equations \eq{2.4} for $\G_\infty$ and $\G_H$ have to include the condition $H_i=0$.

\subsection{Asymptotic charges}

Asymptotic charges are obtained from the variational equation for $\G_\infty$, with $\xi=\pd_t$ (energy) and $\xi=\pd_\vphi$ (angular momentum), using the normalization $16\pi a_0=1$.

To calculate energy, we use the asymptotic formulas $N=1-m/r+O_2$, $(C,D)=O_3$, and the relations
\bea
&&b^0{}_t=N\, ,\hspace{43pt} b^0{}_\vphi=Na\sin^2\th\, ,\hspace{46pt}
  b^1{}_t=b^1{}_\vphi=0\,,                                             \nn\\
&&b^3{}_t=\frac{\sin\th}{\r}a\,,\qquad
  b^3{}_\vphi=\frac{\sin\th}{\r}(r^2+a^2)\,,\qquad
  b^2{}_t=b^2{}_\vphi=0\,,\qquad                                       \nn
\eea
to identify the nonvanishing contributions to $\d\G_\infty[\pd_t]$  (integration implicitly understood)
\bea
&&\d\om^{12}H_{12t}
  =\d\Big(\frac{Nr}{\r^2}b^2\Big)\Big[-2a_0Nb^3\Big]=8\pi a_0\d m\, ,  \nn\\
&&\d\om^{13}H_{13t}
  =\d\Big(\frac{Nr}{\r^2}b^3\Big)2a_0Nb^2=8\pi a_0\d m\, .
\eea
Summing up, one obtains the energy of the Kerr black hole as
\be
E:=\G_\infty[\pd_t]= m\, .
\ee

As for the angular momentum, we use the asymptotic relations
\bea
&&\tom^{01}=
  -\Big(\frac{a}{r}+\frac{ma}{r^2}\Big)\sin^2\th d\vphi+\cO_3\,,   \nn\\
&&\tom^{13}=\Big(1-\frac{m}{r}\Big)\sin\th d\vphi+\cO_2\,,         \nn\\
&&H_{01}=-2a_0(r^2+a^2)\sin\th d\th d\vphi\,,                      \nn\\
&&H_{13}=-2a_0a(r-m)\sin^2\th d\th d\vphi+\cO_1\, ,                \nn
\eea
to calculate the surviving contributions to $\G_\infty[\pd_\vphi]$:
\bea
&&\d\tom^{01}H_{01\vphi}
  =2a_0\big[r\d a+\d(ma)\big]\sin^3\th d\th d\vphi\,,              \nn\\
&&\tom^{13}{_\vphi}\d H_{13}
  =2a_0\big[-r\d a+\d(am)+m\d a\big]\sin^3\th d\th d\vphi\,,       \nn\\
&&\d\tom^{13}H_{13\vphi}=2a_0\big[a\d m\big]\sin^3\th d\th d\vphi\, .
\eea
Summing up these terms and integrating over $S_\infty$, one obtains
\be
\d\G_\infty[\pd_\vphi]
   =12\pi a_0\d(ma)\int_0^{2\pi}\sin^3\th d\th=16\pi a_0\d(ma)\,,\nn\\
\ee
which yields the value for the angular momentum
\be
J:=\G[\pd_\vphi]=ma\, .
\ee

As a consequence, the asymptotic charge associated to $\xi=\pd_t-\Om_+\pd_\vphi$ has the form
\be
\G_\infty[\xi]=m-\Om_+J\, .                                      \lab{4.7}
\ee

\subsection{Entropy and the first law}

Entropy is obtained from the variational equation for $\G_H[\xi]$ with $\xi=\pd_t-\Om_+\pd_\vphi$. Starting with the relations
\bsubeq
\be
\xi\inn b^0\big|_{r_+}=N(1-\Om_+a\sin^2\th)\, ,\qquad
  \xi\inn b^a\big|_{r_+}=0\,,
\ee
where $a=1,2,3$, one obtains
\bea
&&\xi\inn H_{ij}=\xi\inn\om^{Ac}=\xi\inn\om^{23}=0\, ,
  \qquad NN'=\frac{r_+-m}{\r^2}\, ,                                 \nn\\
&&\xi\inn\om^{01}=-NN'(1-\Om_+a\sin^2\th)=-\frac{r_+-m}{r_+^2+a^2}=-\k\,,
\eea
where $A=(0,1),c=(2,3)$, and consequently
\be
\xi\inn\om^{01}\d H_{01}=-\k
  \d\Big[(-2a_0+4b_1C)(r_+^2+a^2)\Big]\sin\th d\th d\vphi\,.
\ee
\esubeq
Then, the calculation of entropy yields
\bea
&&\d\G_H[\xi]=\oint_H(\xi\inn\om^{01})\d H_{01}
             =8\pi a_0\k\d(r_+^2+a^2)\, ,                             \nn\\
\Ra&&\d\G_H[\xi]=T\d S\,,\qquad S=\pi(r_+^2+a^2)\, .              \lab{4.9}
\eea
The result is obtained by noting that the quadratic curvature contributions to $\d H_{01}$ vanish, as follows from
\be
\oint_{S_H}C b^2b^3=2\pi\,,\qquad  \oint_{S_H}Db^2b^3=0\,.        \lab{4.10}
\ee

Combining the previous results with the identity \cite{x10}
\be
\frac{\k}{2}\d(r_+^2+a^2)=\d m-\Om_+\d J\, ,                      \lab{4.11}
\ee
one arrives at the standard GR form of the first law:
\be
\d\G_\infty[\xi]=\d\G_H[\xi]\quad\Lra\quad  \d m-\Om\d J=T\d S\,.
\ee

\subsection{Reduction to GR}

Since the quadratic curvature contributions to the asymptotic charges and entropy effectively vanish, the PG results are directly reduced to the case of GR without any modification. Note that this conclusion differs from the result found by Jacobson and Kang \cite{x13}, obtained in the context of Riemannian $R^2$ theories.

\section{Kerr black hole with torsion}\label{sec5}
\setcounter{equation}{0}

\subsection{Chen et al. solution}\label{sub51}

There exists only one PG solution with Kerr metric and nonvanishing torsion, the solution constructed by Chen et al. \cite{x14}. It is based on a special form of the Lagrangian \eq{2.1}, determined by the following choice of parameters:
\bea
&&(a_1,a_2,a_3)=(a_0,-2a_0,a_3)\,,                                \nn\\
&&b_n=0\text{~~for~~} n\ne 6\,,\qquad \L=0\, .                    \lab{5.1}
\eea
By adopting an ansatz for torsion defined by only one nonvanishing component,
\bsubeq\lab{5.2}
\be
T^0:=f b^0b^1\, ,\qquad f:=f(r,\th)\, ,
\ee
the field equations \eq{2.2} lead to
\be
f(r,\th)=\frac{c_0}{\r\,\sqrt{\D}}\, ,                            \lab{5.2b}
\ee
\esubeq
where $c_0$ is an integration constant.
A direct calculation of the irreducible parts of the field strengths yields $\ir{3}T_i=0$ and $\ir{3}R_{ij}=\ir{6}R_{ij}=0$,
which implies that the Lagrangian parameters $a_3$ and $b_6$ are effectively absent from the field equations.

In a detailed analysis of this solution, McCrea et al. \cite{x15} concluded, among other things, that \emph{nonuniqueness} of the torsion parameter $c_0$ is a serious obstacle to finding a reliable physical interpretation of the solution. A natural criterion for a solution to be unique is based on the possibility to relate its integration constants to certain physical characteristics, like the asymptotic charges, entropy, and so on. However, the situation with Chen et al. solution is more complicated. Namely, for $c_0\ne 0$, the quadratic torsion invariant
\be
T^i\hd T_i=\frac{c_0^2}{\r^2\D}\heps                                \lab{5.3}
\ee
is not only nonunique but also singular. The singularity at $\r^2=0$ is of the same type as the one of the Riemannian curvature \eq{3.9}, but the hypersurface $\D=0$ ($r=r_\pm$), which is a potential location of the horizon, becomes here a true \emph{torsion singularity}. The gravitational singularity without a horizon is known as the naked singularity, where entropy cannot be even defined. Thus, for $c_0\ne 0$, the Chen at al. solution fails to be physically acceptable.

The only way out from this problematic situation is to require $c_0=0$, whereupon the Chen et al. solution reduces to a Riemannian Kerr black hole in PG, and consequently, asymptotic charges and entropy take the GR form, as shown in section \ref{sec4}.

The absence of any physically acceptable Kerr-like solution with $\l=0$ and nonvanishing torsion in the quadratic PG is due to an unusual dynamical mechanism, according to which the curvature squared terms in $L_G$ ``produce" a nonvanishing $\l$, see Ref. \cite{x4}, section 16. The absence of this mechanism in the Chen et al. solution is caused by the fact that the only curvature squared term in $L_G$ is the square of $\irr{6}R^{ij}$, but $\irr{6}R^{ij}$ vanishes on shell.

\subsection{Kerr black hole in teleparallel gravity}\label{sub52}

Teleparallel gravity is a special case of PG, defined by the condition of vanishing Riemann-Cartan curvature, $R^{ij}=0$ \cite{x4}.
Considering the ``inverse" mechanism discussed at the end of the previous subsection, one can conclude that for $\l=0$, a Kerr-AdS solution of the quadratic PG, see for instance \cite{x16,x17}, becomes a Kerr solution with $R^{ij}=0$. This situation motivates us to explore thermodynamic aspects of Kerr spacetime in TG.

The condition $R^{ij}=0$ does not imply that the related ``pure gauge" connection $\om^{ij}$ vanishes. However, since $\om^{ij}$ has no effect on the PG dynamics, see Pereira and Obukhov \cite{x18}, we adopt the simplest choice $\om^{ij}=0$. As a consequence, the tetrad field remains the only dynamical variable, and torsion takes the form $T^i=d b^i$. For Kerr spacetime with tetrad \eq{3.5}, the torsion components are
\bea
&&T^0=-N'b^0b^1+\frac{a^2\sin 2\th}{2\r^3}b^0b^2
      +\frac{2Na\cos\th}{\r^2}b^2b^3\,,                               \nn\\
&&T^1=\frac{a^2\sin 2\th}{2\r^3}b^1b^2\,,\qquad
  T^2=\frac{Nr}{\r^2}b^1b^2\, ,                                       \nn\\
&&T^3=\frac{2ar\sin\th}{\r^3}b^0b^1+\frac{Nr}{\r^2} b^1b^3
       +\frac{(r^2+a^2)\cos\th}{\r^3\sin\th}b^2b^3\,.
\eea
All three irreducible parts of $T^i$ are nonvanishing.

The general (parity invariant) TG Lagrangian is given by
\bsubeq
\be
L_T:=a_0T^i\,\hd\left(a_1\ir{1}T_i+a_2\ir{2}T_i+a_3\ir{3}T_i\right)\,.
\ee
Of particular importance for the physical interpretation of TG is the fact that the choice
\be
(a_1,a_2,a_3)=(1,-2,-1/2)\,,
\ee
\esubeq
defines the theory known as the teleparallel equivalent of GR, \tgr; for more details see \cite{x3,x4}. This equivalence ensures that every vacuum solution of GR is also a solution of \tgr. In particular, this is true for Kerr spacetime. In spite of such a dynamical equivalence, the geometric content of the two theories is quite different: GR is characterized by a Riemannian curvature and vanishing torsion, whereas the teleparallel geometry of \tgr\ has a nontrivial torsion but vanishing curvature.

Thermodynamic properties of Kerr black hole in \tgr\ are determined by the covariant momentum,
\bsubeq
\be
H^i=2a_0\,\hd\Big(\ir{1}T^i-2\ir{2}T^i-\frac{1}{2}\ir{3}T^i\Big)\,,
\ee
the explicit form of which reads
\bea
H^0&=&2a_0\left[\frac{aN\cos\th}{\r^2}b^0b^1
      -\frac{ar\sin\th}{\r^3}b^0b^2+\frac{\cos\th}{\r\sin\th}b^1b^3
      -\frac{2Nr}{\r^2}b^2b^3\right]\,,                                \nn\\
H^1&=& 2a_0\left[\frac{\cos\th}{\r\sin\th}b^0b^3
             -\frac{ar\sin\th}{\r^3}b^1b^2\right]\,,                   \nn\\
H^2&=&2a_0\left[-\frac{r-m}{N\r^2}b^0b^3
               +\frac{Na\cos\th}{\r^2}b^1b^2\right]\,,                 \nn\\
H^3&=&2a_0\left[\frac{a^2\sin 2\th}{\r^3}b^0b^1+\frac{r-m}{N\r^2}b^0b^2
      +\frac{Na\cos\th}{\r^2}b^1b^3          -\frac{ar\sin\th}{\r^3}b^2b^3\right]\,.\quad
\eea
\esubeq

\subsubsection{Asymptotic charges}

In the calculation of $\d\G_\infty[\xi]$, the variational operator $\d$ should not act on the background configuration, as specified in section \ref{sec2}, rule a1). For the Kerr metric, the background configuration is defined by $m=0$, and it has the form of the Minkowski spacetime $M_4$, described in terms of some unusual coordinates that depend on the parameter $a$ \cite{x10}. Hence, the variation $\d$ should not be applied to those $a$'s that correspond to $M_4$. The simplest way to realize this requirement is by using an equivalent procedure: first formally apply $\d$ to all $a$'s, then remove $m$-independent terms, stemming from the variation of $M_4$.

The energy of the teleparallel Kerr solution is obtained from just one nonvanishing term:
\bsubeq
\bea
&&\d\G_\infty=\oint_{S_\infty} b^0{_t}\d H_0=16\pi a_0\d m \,,         \\
\Ra&& E:=\G_\infty=m\, .
\eea
\esubeq

Concerning the angular momentum, the formal variation over $a$ and $m$ yields the following nonvanishing terms:
\bea\lab{6.5}
&&\oint_{S_\infty} b^0{_\vphi}\d H_0=\frac{32\pi a_0}{3}a\d m\,,\nn\\
&&\oint_{S_{\infty}}\d b^0 H_{0\vphi}
          =\frac{32\pi a_0}{3}\left[-\ul{r_{\infty}\d a}_\times
                                    +\d(am)+m\d a\right]\, ,           \nn\\
&&\oint_{S_\infty} b^3{_\vphi}\d H_3
    =\frac{16\pi a_0}{3}\left[2\ul{r_{\infty}\d a}_\times-\d(am)\right].
\eea
Here, the underlined terms are divergent (but their sum formally vanishes). However, since they are $m$-independent and consequently, associated to the variation of $M_4$, their contribution should be simply omitted. The sum of the remaining terms yields the finite angular momentum:
\bsubeq
\bea
&&\d\G_\infty[\pd_\vphi]=16\pi a_0 \d(am)\, ,                           \\
\Ra&& J:=\G_\infty[\pd_\vphi]=ma\,.
\eea
\esubeq

Thus, the conserved charge defined by the Killing vector $\xi=\pd_t-\Om_+\pd_\vphi$ reads
\be
\G_\infty[\xi]=m-\Om_+J\,.
\ee

\subsubsection{Entropy and the first law}

The entropy of the teleparallel Kerr black hole can be computed from the
following surviving contributions to $\d\G_H$:
\bea
&&\d b^2(\xi\inn H_2)=2a_0\frac{r_+-m}{2mr_+}(\d b^2)b^3\,,\nn\\
&&\d b^3(\xi\inn H_3)=2a_0\frac{r_+-m}{2mr_+}b^2 (\d b^3)\,,
\eea
where we used
$$
\frac{1-\Om_+a\sin^2\th}{r_+^2+a^2\cos^2\th}=\frac{1}{2m r_+}\,.
$$
As a consequence,
\bsubeq
\bea
&&\d\G_H=2a_0\k\oint\d(b^2b^3)=\frac{\k}{2}\d(r_+^2+a^2)\,.           \\
\Ra&&\d\G_H=T\d S\, ,\qquad S=\pi (r_+^2+a^2)\,.
\eea
\esubeq

The above results, combined with the identity \eq{4.11}, imply the validity of the first law:
\be
T\d S=\d m-\Om_+\d J\,.
\ee

\section{Concluding remarks}\label{sec6}
\setcounter{equation}{0}

In this paper, we used the Hamiltonian approach developed in Ref. \cite{x6} to investigate entropy and the first law of black hole thermodynamics of Kerr-like spacetimes. Analyzing two possible cases, we found the results that coincide with those in GR.

Case A: Kerr spacetime can be understood as a Riemannian solution of PG. In that case, the contribution of the quadratic curvature term in $L_G$ to entropy effectively vanishes. Hence, entropy is determined as in GR---by the only remaining term, the scalar curvature.

Case B: Kerr black hole with torsion is treated as a solution with Kerr metric and nonvanishing torsion in PG/\tgr. First, we analyzed the Chen et al. solution of PG \cite{x14}. The singularity of its quadratic torsion invariant is shown to be a naked singularity, which can be avoided only by choosing the torsion function to vanish. Thus, it is only the Riemannian version of the solution that has an acceptable physical interpretation. Second, we studied Kerr spacetime as a solution of \tgr, where torsion is the only nonvanishing field strength. The Hamiltonian approach offers a simple dynamical mechanism to explain how Cartan's torsion can produce an amount of black hole entropy that is exactly the same as in GR. In view of the dynamical equivalence of GR and \tgr, this result, which is not a surprise, represents a consistency test of the present approach to entropy.

The entropy of the Kerr-like solutions follows the pattern of spherically symmetric solutions studied in Ref. \cite{x6}: up to a possible multiplicative constant, stemming from the presence of the quadratic field strengths in the PG Lagrangian, black hole entropy and asymptotic charges are of the standard GR form. Why is that so? In an attempt to better understand the dynamical meaning of these results, we noticed an interesting counterexample, the Banados-Teitelboim-Zanelli black hole with torsion in three-dimensional gravity \cite{x19}. The Lagrangian is constructed not only from the field strengths but includes also a Chern-Simons term. It is exactly this parity-violating term that modifies asymptotic charges and entropy with respect to their GR values: these quantities become linear combinations of the integration parameters $m$ and $J$, but the first law remains valid. Such a situation naturally suggests to explore possible influence of the parity-violating terms in PG, see for instance \cite{x20,x17}, on the form of black hole entropy.

Since genuine exact solutions of quadratic PG are characterized by an effective cosmological constant, our future research of black hole entropy will be focussed on the class of Kerr-(Anti)-de Sitter solutions in PG.

\section*{Acknowledgments}

We wish to thank Friedrich Hehl and Yuri Obukhov for critical comments on the third (published) version of the manuscript. This work was partially supported by the Serbian Science Foundation under Grant No. 171031. The results are checked using the computer algebra system {\sc Reduce}.

\vsm


\begin{thebibliography}{99}

\bibitem{x1} R. M. Wald, Black hole entropy is the Noether charge, Phys. Rev. D {\bf 48}, R3427 (1993).

\bibitem{x2} R. M. Wald, The thermodynamics of black holes, Living Rev. in Rel. {\bf 4}, 6 (2001) (44 pages) [gr-qc/9912119].

\bibitem{x3} M. Blagojevi\'c, \emph{Gravitation and Gauge Symmetries} (IoP, Bristol, 2002).

\bibitem{x4}   M. Blagojevi\'c and F. W. Hehl (eds.), \emph{Gauge Theories of Gravitation, A Reader with Commentaries} (Imperial College Press, London, 2013).

\bibitem{x5} Yu. N. Obukhov, Poincar\'e gauge gravity: Selected topics, Int. J. Geom. Methods Mod. Phys. {\bf 03} (2006) 95-138.

\bibitem{x6} M. Blagojevi\'c and B. Cvetkovi\'c, Entropy in Poincar\'e gauge theory: Hamiltonian approach, Phys. Rev. D {\bf 99} (2019) 104058 (12 pages).

\bibitem{x7} R. P. Kerr, Gravitational field of a spinning mass as an example of algebraically special metrics, Phys. Rev. Lett. {\bf 11}  (1963) 237-238;

    R. H. Boyer and R. W. Lindquist, Maximal analytic extension of the Kerr metric, J. Math. Phys. {\bf 8} (1967) 265-281.

\bibitem{x8} S. Chandrasekhar, \emph{The mathematical theory of black holes} (Oxford University Press, Oxford, 1983);

    M. Visser, The Kerr spacetime: A brief introduction, in: \emph{The Kerr Spacetimes, Rotating Black holes in General Relativity}, edited by D. L. Wiltshire, M. Visser, S. M. Scott (Cambridge University Press, Cambridge, 2009), pp. 3-37, arXiv:0706.0622 [gr-qc].

\bibitem{x9} Barret O'Neill, \emph{The geometry of Kerr black holes} (A. K. Peters Ltd., Wellesley Massachusetts, 1995).

\bibitem{x10} E. Poisson, \emph{A Relativist's Toolkit, The Mathematics of Black-Hole Mechanics} (Cambridge University Press, Cambridge, England, 2004).

\bibitem{x11} J. B. Griffiths and J. Podolsky, \emph{Exact Space-Times in Einstein's General Relativity} (Cambridge University Press, Cambridge,
    England, 2009).

\bibitem{x12} Ch. Heinicke and F. W. Hehl, Schwarzschild and Kerr solutions
     of Einstein's field equation, an introduction, Int. J. Mod. Phys. D
     {\bf 24} (2015) 1530006 (78 pages).

\bibitem{x13} T. Jacobson and G. Kang, On black hole entropy, Phys. Rev. D  {\bf 49} (1994) 6587-6598.

\bibitem{x14} H. -H. Chen, D. -C. Chern, R. -R. Hsu, W. B. Yeung, and M. -Q. Chen, An axisymmetric solution to the Poincar\'e gauge theory of gravitation with Kerr metric and nonvanishing torsion, Chin. J. Phys. (Taiwan) {\bf 24} (1986) 115-121.

\bibitem{x15} J. D. McCrea, E. W. Mielke, and F. W. Hehl, A remark on the axisymmetric Chen et al. solution of the Poincar\'e gauge theory,
    Phys. Lett A {\bf 127} (1988) 65-69.

\bibitem{x16} J. D. McCrea, P. Baekler and M. G\"urses, A Kerr-like solution of the Poincar\'e gauge field equations, Nuovo Cimento B {\bf 99} (1987) 171-177;

    P. Baekler, M. G\"urses, F. W. Hehl, and J. D. McCrea, The exterior gravitational field of a charged spinning source in the Poincar\'e gauge theory: A Kerr-Newman metric with dynamic torsion, Phys. Lett. A
    {\bf 128} (1988) 245-250.

\bibitem{x17} Yu. N. Obukhov, Exact solutions in Poincar\'e gauge gravity theory, Universe {\bf 5, 127} (2019) (13 pages).

\bibitem{x18} J. G. Pereira and Yu. N. Obukhov, Gauge structure of teleparallel gravity, Proceedings of the workshop \emph{Teleparallel Universes in Salamanca}, Salamanca, Spain, 26-28 November 2018.

\bibitem{x19} M. Blagojevi\'c and B. Cvetkovi\'c, Black hole entropy in 3D    gravity with torsion, Class. Quantum Gravity {\bf 23} (2006) 4781-4795.

\bibitem{x20} G. K. Karananas, The particle spectrum of parity-violating
    Poincar\'e gravitational theory, Class. Quantum Gravity {\bf 32}
    (2015) 055012 (38 pages);

    M. Blagojevi\'c and B. Cvetkovi\'c, General Poincar\'e gauge theory: Hamiltonian structure and particle spectrum, Phys. Rev. D {\bf 98} (2018) 024014 (21 pages);



\end{thebibliography}
\end{document}